\documentclass[11pt]{article}
\usepackage{graphicx}
\begin{document}
%

\begin{center}
{\large \bf Neutrino Oscillations as a Lepton-Flavor-Violating Interaction}

\vskip.5cm

W-Y. Pauchy Hwang\footnote{Correspondence Author;
 Email: wyhwang@phys.ntu.edu.tw; The second version of
 arXiv:1207.6443v1 [hep-ph] 27 Jul 2012.} \\
{\em Asia Pacific Organization for Cosmology and Particle Astrophysics, \\
Institute of Astrophysics, Center for Theoretical Sciences,\\
and Department of Physics, National Taiwan University,
     Taipei 106, Taiwan}
\vskip.2cm


{\small(24 July 2012; revised: August 15, 2013)}
\end{center}

\begin{abstract}
To describe neutrino oscillations in the sense of quantum mechanics
and quantum field theory, we propose to use an off-diagonal
neutrino-Higgs (mass) interaction, as discussed originally in a
family gauge theory and in the extended Standard Model. For neutrino
oscillations which take place presumably between point-like Dirac
particles, the proposed description would be unique in the
quantum mechanics sense. This may help us to resolve
a few outstanding puzzles - the question of why there are only
three generations, the question of why the masses of neutrinos
are so tiny, the question of why neutrinos oscillate, and the
question of why the dark-matter world is so huge (25\%) as
compared to the visible ordinary-matter world (5\%).

\bigskip

{\parindent=0pt PACS Indices: 12.60.-i (Models beyond the standard
model); 98.80.Bp (Origin and formation of the Universe); 12.10.-g
(Unified field theories and models).}
\end{abstract}

\bigskip

\section{Why are neutrinos so interesting?}

\medskip

Neutrinos have masses, the tiny masses far below the range of the masses of
the quarks and charged leptons. Maybe due to the non-zero masses, neutrinos
oscillate among themselves, giving rise to a lepton-flavor violation (LFV).
Neutrino masses and neutrino oscillations may be regarded as one of
the most important experimental facts over the last thirty years \cite{PDG}.

Certain LFV processes such as $\mu \to e + \gamma$ \cite{PDG}
and $\mu + A \to A* + e$ are closely related to the most cited picture of
neutrino oscillations so far \cite{PDG} - they also occur, however tiny,
if neutrinos oscillate. In this note, I wish to point
out that the cross-generation or off-diagonal neutrino-Higgs interaction
may serve as the detailed mechanism of neutrino oscillations, with some
vacuum expectation value (VEV) of the new Higgs field(s).

Presumably just like other building blocks of matter such as quarks and charged
leptons, we could treat these neutrinos as point-like Dirac particles. Then,
neutrino oscillations are fundamental and deep, certainly deeper than oscillations
in other composite systems - such as oscillations in the $K^0-\bar K^0$ system.
Thus, it would be natural to describe the reaction as $i \eta \bar \Psi \times
\Psi \cdot \Phi$ with some VEV for the family Higgs field $\Phi$, where $\bar \Psi$
and $\Psi$ are family-triplet Dirac fields. Here the curl-dot product is to be
explained later. The existence of this unique story for neutrino oscillations
is amazing.

For over half a century, we have the outstanding question why there are
three generations in the minimal Standard Model \cite{Books}. And, for
the last decade, another outstanding puzzle emerges that the dark-matter
world is about five times the visual ordinary-matter world (the latter
described by the minimal Standard Model). Besides the role in the minimal
Standard Model, neutrinos may be able to tell us something in the
dark-matter world which our neutrinos are capable of talking to (or
interacting with).

Indeed, there is room left for something very interesting. Remember that
the right-handed neutrinos never enter in the construction of the minimal
Standard Model \cite{Books}. The message that the right-handed neutrinos
seem to be "unwanted" could be telling us something. Now, the fact that
neutrinos have tiny masses suggests that "more naturally" they would be
four-component Dirac particles, and unlikely to be the two-component
Majorana particles.

The room left for the right-handed neutrinos is that they are "unwanted"
in the minimal Standard Model and that they could form some multiplet(s)
under a new (dark-matter) gauge group besides the minimal Standard Model.
We have some candidate from the symmetries - the family symmetry that
there are three generations in the building blocks of (ordinary) matter,
and so far only three. The puzzle so well-known that we no longer question
ourselves why or why not! We have seen this fact, but we don't know why
- let's speculate that it could be the story associated with the
dark-matter world.

It arises naturally the so-called family gauge theory \cite{Family}. Note
that the right-handed neutrinos do not appear in the minimal Standard Model.
So, we could make a massive $SU_f(3)$ gauge theory completely independent of
the minimal Standard Model, including the particle content. We
could treat $(\nu_{\tau R},\, \nu_{\mu R},\, \nu_{e R})$ as a triplet under
this $SU_f(3)$ - so to give rise to a family gauge theory. This completes
the derivation of the family gauge theory \cite{Family}. The $SU_f(3)$ is
by definition the massive gauge theory - all the involved particles, except
the neutrinos, are massive dark-matter particles.

\medskip

\section{Neutrino Oscillations as a Lepton-Flavor-Violating Interaction}

So, the question becomes: Can we construct the overall consistent theory
based on the group $SU_c(3) \times SU_L(2) \times U(1) \times SU_f(3)$,
i.e., to add an extra $SU_f(3)$ to the minimal Standard Model?

The answer is an amazing "yes". The first step is to decide what our
"basic units" (out of the building blocks of matter) are and how many
they are. For instance, the right-handed neutrino triplet
$(\nu_{\tau R},\, \nu_{\mu R},\, \nu_{e R})$ ($\equiv
\Psi_R (3,1)$ - $SU_f(3)$ triplet and $SU_L(2)$ singlet) would be a
"basic unit". In Hwang and Yan \cite{HwangYan}, we assign three
$SU_f(3)$ fermion triplets - $\Psi_L(3,2)$, $\Psi_R(3,1)$, and
$\Psi_R^C(3,1)$ (charged). All quarks are singlets under $SU_f(3)$.
As the major second step, we have to check whether the complicated
Higgs mechanisms would work out. This is the "amazing" part of the
story. In the extended Standard Model \cite{Hwang417}, we have three
scalar/Higgs fields: the Standard-Model Higgs $\Phi (1,2)$, the
family Higgs triplet $\Phi(3,1)$, and the mixed family-triplet
and $SU_L(2)$-doublet scalar/Higgs $\Phi(3,2)$. In the U-gauge,
the Standard-Model Higgs picks out the neutral component (one degree
of freedom), which in turn projects out the neutral components in
$\Phi(3,2)$ such that the neutral part has the spontaneous symmetry
breaking (i.e. the "projected-out Higgs mechanism") but the charged
part remains to be massive. The neutral part of $\Phi(3,2)$ and
the family Higgs $\Phi(3,1)$ make the eight gauge bosons very
massive, leaving four real family Higgs.

Using this language \cite{Hwang417}, we can write the mass term of the
neutrinos:

\begin{equation}
i \eta {\bar \Psi}_L(3,2) \times \Phi(3,2) \cdot \Psi_R(3,1) + h.c.,
\end{equation}
which is an off-diagonal matrix (in $SU_f(3)$). Although it is trivial, the
operation does not belong to the mathematics of the matrix. That is,
$\nu_e$ would transform into $\nu_\mu$ or into $\nu_\tau$, $\nu_\mu$ would into
$\nu_\tau$ or $\nu_e$, and so on. This is interesting in view of neutrino
oscillations, since it could be regarded as the underlying interaction
(mechanism) for neutrino oscillations (which we are talking about \cite{PDG}).
An oscillation occurs in a way similar to the decay by way of creating a new
species plus the vacuum expectation value (or, changing the vacuum). In quantum
mechanics, this may be so far the only way how an oscillation can occur.

To illustrate the point further, we calculate the golden lepton-flavor-violating
decay $\mu \to e + \gamma$ as the celebrated example. We show in Figures 1(a),
1(b), and 1(c) three leading basic Feynman diagrams. Here the conversion
of $\nu_\mu$ into $\nu_e$ is marked by a cross sign and it is a term from our
off-diagonal interaction given above with the Higgs vacuum expectation values
$u_+$ and $u_-$. Here the Higgs masses are assumed to be very large, i.e.,
greater than a few $TeV$, as in $SU_f(3)$. The only small number (coupling) is
$\eta$, explaining the tiny masses of neutrinos.

\begin{figure}[h]
\centering
\includegraphics[width=4in]{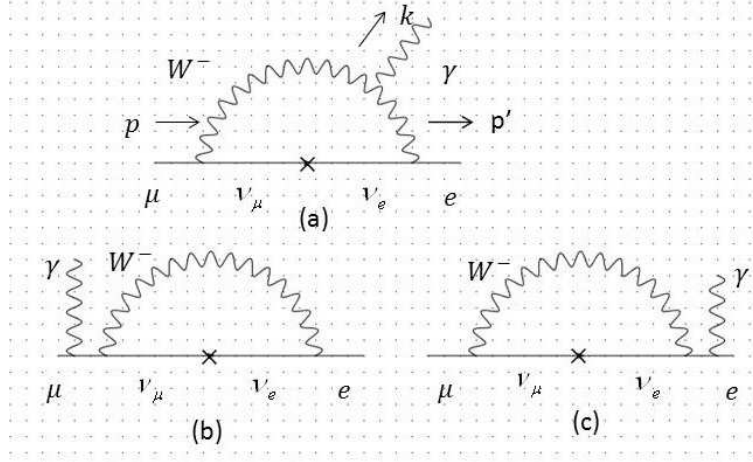}
\caption{The leading diagrams for $\mu \to e + \gamma$.}
\end{figure}

Using Feynman rules from Wu and Hwang \cite{Books}, we write, for Fig. 1(a),

\begin{eqnarray*}
{1\over (2\pi)^4} \int d^4q \cdot {\bar u}(p',s')\cdot &i \cdot {1\over 2 \sqrt 2}
{e\over sin \theta_W}\cdot i \gamma_\lambda (1+ \gamma_5)\nonumber\\
\cdot {1\over i} {m_2-i\gamma\cdot q\over {m_2^2+q^2-i\epsilon}}\cdot
&i \cdot i \eta (-) u(\nu_\mu\to \nu_e) \cdot {1\over i} {m_1-i\gamma\cdot q\over
{m_1^2 + q^2-i\epsilon}} \nonumber\\
\cdot i\cdot {1\over 2 \sqrt 2}{e\over sin\theta_W}\cdot &i \gamma_{\lambda'}
(1+\gamma_5)\cdot u(p,s)\nonumber\\
\cdot {1\over i} {\delta_{\lambda'\mu}\over {M_W^2+(p-q)^2-i\epsilon}}\cdot
{\epsilon_\sigma(k)\over \sqrt{2k_0}}\cdot &\Delta_{\sigma\mu\nu} \cdot
{1\over i} {\delta_{\nu\lambda}\over {M_W^2+(p-q-k)^2-i\epsilon}},
\end{eqnarray*}
with $\Delta_{\sigma\mu\nu}=(-ie)\{\delta_{\mu\nu}(-k-p-q)_\sigma +
\delta_{\nu\sigma}(p-q+p-q-k)_\mu +\delta_{\sigma\mu} (-p+q+k+k)_\nu \}$.

On the other hand, Feynman rules yield, for Fig. 1(b),
\begin{eqnarray*}
{1\over (2\pi)^4} \int d^4q \cdot {\bar u}(p',s')\cdot &i \cdot {1\over 2 \sqrt 2}
{e\over sin \theta_W}\cdot i \gamma_\lambda (1+ \gamma_5)\nonumber\\
\cdot {1\over i} {m_2-i\gamma\cdot q\over {m_2^2+q^2-i\epsilon}}\cdot
&i\cdot i \eta (-)u(\nu_\mu\to \nu_e)\cdot {1\over i} {m_1-i\gamma\cdot q\over
{m_1^2 + q^2-i\epsilon}} \nonumber\\
\cdot i\cdot {1\over 2 \sqrt 2}{e\over sin\theta_W}\cdot &i \gamma_{\lambda'}
(1+\gamma_5)\cdot \nonumber\\
\cdot {1\over i} {\delta_{\lambda\lambda'}\over {M_W^2+(p'-q)^2-i\epsilon}}
\cdot {1\over i} {m_\mu - 1\gamma\cdot p'\over {m_\mu^2+ p^{\prime 2}-i\epsilon}}
\cdot &i (-i)e \gamma_\sigma\cdot {\epsilon(k)\over \sqrt {2k_0}}. u(p,s),
\end{eqnarray*}
and a similar result for Fig. 1(c).

The four-dimensional integrations can be carried out, via the dimensional
integration formulae (e.g. Ch. 10, Wu/Hwang \cite{Books}), especially
if we drop the small masses compared to the W-boson mass $M_W$ in the
denominator. In this way, we obtain
\begin{eqnarray*}
i T_a={G_F\over \sqrt 2} &\cdot \eta u(\nu_\mu\to \nu_e)
\cdot (m_1 + m_2)\cdot (-2i){e\over (4\pi)^2}\nonumber\\
&\cdot {\bar u}(p',s') {\gamma\cdot \epsilon\over \sqrt {2k_0}}
(1+\gamma_5) u(p,s).
\end{eqnarray*}

It is interesting to note that the wave-function renormalization,
as shown by Figs. 1(b) and 1(c), yields
\begin{eqnarray*}
i T_{b+c}= {G_F\over \sqrt 2} &\cdot \eta u(\nu_\mu\to \nu_e) (m_1 + m_2)
\cdot (+2i){e\over (4\pi)^2}
\cdot \{{p'^2\over m_\mu^2 + p'^2} + {p^2\over m_e^2 + p^2}\}\nonumber\\ &\cdot
{\bar u}(p',s') {\gamma \cdot \epsilon \over \sqrt{2k_0}} (1+\gamma_5) u(p,s),
\end{eqnarray*}
noting that $p^2=-m_\mu^2$ and $p'^2=-m_e^2$ would make the contribution
from Figs. 1(b) and 1(c) to be the same as, but opposite in sign, that
from Fig. 1(a). This is a manifestation of "gauge invariance".

In a normal treatment, one ignores the wave-function renormalization
diagrams 1(b) and 1(c) in the treatment of the decays $\mu \to e + \gamma$,
$\mu \to 3e$, and $\mu+ A \to e+ A^*$. Thus, one may ignore some important
cancelation, if any.

Comparing this to the dominant mode $\mu \to e {\bar \nu}_e \nu_\mu$
\cite{Books}, we could obtain the branching ratio.
The decay rate for $\mu \to e+ \gamma$, as would be obtained here,
would be of the order $(m_{neutrino}\cdot m_\mu)^2/M_W^4$, which is
extremely small.

The off-diagonal mass matrix would be modified by the self-energy
diagram since the neutrinos form a triplet under $SU_f(3)$. It is
presumed that these self-energy diagrams, after the ultraviolet
divergences get subtracted, lead to masses of the right order.

The four family Higgs have to belong to two triplets - the neutral
part of $\Phi(3,2)$ and the purely family Higgs $\Phi(3,1)$. If
it is two-two divided such as that addressed in \cite{Family},
then the situation would be as follows: If
the off-diagonal mass matrix is diagonalized alone, the three roots
would be two negative and one positive, adding up to zero. This seems
like one ordering in the masses of neutrinos - one up and two downs.
Of course, it could be three-one divided as well.

Besides the golden decay $\mu \to e+ \gamma$ (much too small) and
neutrino oscillations (already observed), violation of the
$\tau-\mu-e$ universality is also anticipated and might be observed.
As the baryon-antibaryon asymmetry is sometime attributed to the
lepton-antilepton asymmetry, the current scenario for neutrino
oscillations \cite{PDG} seems inadequate for this purpose. If we
take the hints from neutrinos rather seriously, there are so much
to discover, even though the minimal Standard Model for the
ordinary-matter world would, by and large, remain to be intact.

Of course, the Standard-Model Higgs has now been discovered. The
direct search for the family Higgs and the massive family bosons
in the $TeV$ range would be too costly. So, the searches for some
lepton-flavor-violating decays and for violation of the
$\tau-\mu-e$ universality would be alternative for the moment.

To sum up, if we treat neutrinos as "point-like" Dirac particles,
the curl-dot product as in Eq. (1) would be the way to go. It is the
way to connect neutrino oscillations with the lepton-flavor-violating
decays or reactions. The curl-dot products are {\it not} the matrix
operations (in the mathematics sense); it represents a new way to
introduce renormalizable interactions and so expands the horizon of
quantum field theory. 

\medskip

\section{Further Thoughts}

We believe that, in the dark-matter world, the dark-matter particles
are also species in the extended Standard Model. Most of reactions
happening among dark-matter particles, even involving neutrinos,
cannot be detected in the ordinary-matter world. It is clear that
the minimum extended Standard Model would be the extended Standard
Model to be based on the group $SU_c(3) \times SU_L(2) \times U(1)
\times SU_f(3)$, since this model is rather unique (and economical).
The issue is whether our Standard Model would close up our world - that 
is, all particles in our world are accounted for. 

In a slightly different context \cite{Hwang3}, I propose to work with
two working rules: "Dirac similarity principle", based on eighty years of
experience, and "minimum Higgs hypothesis", from the last forty years
of experience. Using these two working rules, the extended model mentioned
above becomes rather unique - so, it is so much easier to check it against
the experiments. The close-up question of our world may have to be answered,
provided that the two working rules, or similar, are assumed.   

We would be curious about how the dark-matter world looks like, though
it is difficult to verify experimentally. The first question would be: The
dark-matter world, 25 \% of the current Universe (in comparison, only 5 \%
in the ordinary matter), would clusterize to form the dark-matter
galaxies, maybe even before the ordinary-matter galaxies. The dark-matter
galaxies would then play the hosts of (visible) ordinary-matter galaxies,
like our own galaxy, the Milky Way. Note that a dark-matter galaxy is
by our definition a galaxy that does not possess any ordinary strong and
electromagnetic interactions (with our visible ordinary-matter world).
This fundamental question deserves some thoughts, for the structural
formation of our Universe.

Of course, we should remind ourselves that, in our ordinary-matter world,
those quarks can aggregate in no time, to hadrons, including nuclei, and
the electrons serve to neutralize the charges also
in no time. Then atoms, molecules, complex molecules, and so on. These serve as
the seeds for the clusters, and then stars, and then galaxies, maybe in a time span
of $1\, Gyr$ (i.e., the age of our young Universe). The aggregation caused by
strong and electromagnetic forces is fast enough to help giving rise to galaxies
in a time span of $1\, Gyr$. On the other hand, the seeded clusterings might
proceed with abundance of extra-heavy dark-matter particles such as familons
and family Higgs, all greater than a few $TeV$ and with relatively long
lifetimes (owing to very limited decay channels). So, further simulations on
galactic formation and evolution may yield clues on our problem.

Finally, coming back to the fronts of particle physics, neutrinos might
couple to the dark-matter particles. Any further investigation along this
direction would be of utmost importance. It may shed light on the nature
of the dark-matter world and, eventually, we would be able to close up
our world.

\bigskip

This research is supported in part by National Science Council project (NSC
99-2112-M-002-009-MY3). We wish to thank the authors of the following
books \cite{Books} for thorough reviews of the minimal Standard Model.

\end{document}